\documentclass{PoS}
\usepackage{amsmath}
\usepackage{subfig}

\title{Eigenspectrum Noise Subtraction Methods in Lattice QCD}

\ShortTitle{Eigenspectrum Noise Subtraction Methods in Lattice QCD}

\author{\speaker{Victor Guerrero} \\
        Department of Physics, Baylor University, Waco, TX 76798-7316\\
        E-mail: \email{victor\_guerrero@baylor.edu}}
\author{Ronald B. Morgan \\
Department of Mathematics, Baylor University, Waco, TX 76798-7328, USA \\
E-mail: \email{ronald\_morgan@baylor.edu}}
\author{Walter Wilcox \\
Department of Physics, Baylor University, Waco, TX 76798-7316, USA \\
E-mail: \email{walter\_wilcox@baylor.edu}}

\abstract{We propose a new noise subtraction method, which we call "eigenspectrum subtraction", which uses low eigenmode information to suppress statistical noise at low quark mass. This is useful for lattice calculations involving disconnected loops or all-to-all propagators. It has significant advantages over perturbative subtraction methods. We compare unsubtracted, eigenspectrum and perturbative error bar results for the scalar operator on a small Wilson QCD matrix.}

\FullConference{The XXVII International Symposium on Lattice Field Theory\\
                 July 26-31, 2009\\
                 Peking University, Beijing, China}

\begin{document}

\section{Introduction}
One of the hardest problems in Lattice QCD is the calculation of disconnected quark loops and all-to-all quark propagators \cite{wilcox99}. Statistical noise methods can, in principle, make any matrix element available, but the computer expense can be prohibitive. This is especially true at low quark mass, where error bars are greatest. Noise subtraction methods, such as the  perturbative subtraction method \cite{thron98}, which improve the signal by a suppression of off-diagonal noise, are essential for efficiency. However, perturbative subtraction is ineffective at low quark mass. We propose a subtraction method, which we call "eigenspectrum subtraction", which uses low eigenmode information to suppress the statistical noise at low quark mass. Such eigenmode information is readily available from fermion deflation algorithms, such as GMRES-DR \cite{darnell08} for non-hermitian systems or LAN-DR \cite{rehim} for hermitian ones. 

\section{Method}
\subsection{Introduction}
Our noise methods will be utilizing real Z(2) noise. These are vectors made up of statistically random $1$'s and $-1$'s. A useful property of these Z(2) noise vectors is

\begin{equation}
\delta_{ij}= \lim_{N\to\infty} \frac{1}{N} \sum_{n}^{N}{z_{2_i}^{(n)} z_{2_j}^{(n)}} \equiv \left< z_{2_i} z_{2_j} \right>
\end{equation}

The quark propagator problem can be formed by solving for the solution vector of a linear system made with the necessary action, $M$, and any given source vector, $b$. 

\begin{equation}
\begin{split}
M x &= b \\
\Rightarrow x &= M^{-1} b
\end{split}
\end{equation}

For our noise algorithms we will be using many Z(2) noises as our source vectors so we get

\begin{equation}
x_i^{(n)}=\sum_{j}{M_{i j}^{-1} z_{2_j}^{(n)}}
\end{equation}

This formulation is done with a Wilson action of the form

\begin{equation}
M = 1 - \kappa D
\end{equation}

For the disconnected quark loops and all-to-all quark propagators we will, as a test, form $\text{Tr}(M^{-1})$, although the method can be used for any quark matrix element.

\subsection{Non-Subtraction}
First let us consider the the basic setup. By utilizing the aforementioned properties we can compute the trace of the inverse of a matrix.

\begin{equation}
\begin{split}
\text{Tr}(M^{-1})&=\sum_{j}{M_{j j}^{-1}} \\
                  &=\sum_{i,j}{M_{i j}^{-1}\delta_{ij}} \\
                  &=\sum_{i,j}{M_{i j}^{-1} \left< z_{2_i} z_{2_j} \right>} \\
                  &= \sum_{i,j}\lim_{N\to\infty} \frac{1}{N} \sum_{n}^{N} z_{2_i}^{(n)}{M_{i j}^{-1} z_{2_j}^{(n)}} \\
                  &= \sum_{i}\lim_{N\to\infty} \frac{1}{N} \sum_{n}^{N} z_{2_i}^{(n)} x_i^{(n)} 
\end{split}
\end{equation}

This method allows for the trace computation to be done by solving many linear systems. The error from this method is introduced by the off-diagonal elements, that is; the less diagonally dominant the matrix is, the greater the error that is introduced. In order to minimize this effect we must form a matrix that approximates the off-diagonal elements of $M^{-1}$. We will call such a matrix $\tilde{M}^{-1}$. We will use this matrix to subtract out the off-diagonal elements. 

\subsection{Perturbative Subtraction}
A useful method is to use the perturbative approach to form the approximation matrix\cite{thron98}. That is to form 

\begin{equation}
\tilde{M}_{pert}^{-1} \equiv 1+\kappa D+(\kappa D)^2+(\kappa D)^3+(\kappa D)^4
\end{equation}

Since $M$ is in a form that is conducive to small parameter expansions we can subtract off the necessary, approximate, off-diagonal elements. Since $\tilde{M}_{pert}^{-1}$ is only an approximation it does have on-diagonal elements and thus the trace of $\tilde{M}_{pert}^{-1}$ must be added back. 

\begin{equation}
\text{Tr}(M^{-1})=\lim_{N\to\infty} \frac{1}{N} \sum_{n}^{N} z_{2_i}^{(n)} \sum_{j}{(M_{i j}^{-1}-\tilde{M}_{pert}^{-1}) z_{2_j}^{(n)}}+\text{Tr}\left(\tilde{M}_{pert}^{-1}\right)
\end{equation}

This method produces significant improvements to the ``Non-Subtraction'' method, but with a catch. Since $\tilde{M}_{pert}^{-1}$ is a small parameter expansion, it is most effective for small values of $k$, which relate to ``large'' quark mass values.

\subsection{Eigenspectrum Subtraction}
In order to allow for small values of the hopping parameter we must formulate a new approximation matrix. Our new method allows for an eigenspectrum formulation by utilizing the eigenvectors, both left and right, from the non-hermitian Wilson matrix, $M$. 

Right eigenvectors are formed as

\begin{equation}
 M e_R^{(q)} = \lambda^{(q)} e_R^{(q)}
\end{equation}

where left eigenvectors are formed like so

\begin{equation}
e_L^{(q)^{T}} M = e_L^{(q)^{T}} \lambda^{(q)}
\end{equation}

It is useful to note that the eigenvalues of the left and right eigen-system problem are the same. That is to say for every right eigenvector, there exists a left eigenvector. Left and right eigenvectors are orthogonal.  (For the Wilson case, the two are simply related by $\gamma_5$.) Thus any matrix, $m \times m$ can be formed via an eienspectrum formulation by

\begin{equation}
M = \sum_{q=1}^{m}{e_{R}^{(q)} \lambda^{(q)} e_{L}^{(q)^T}}
\end{equation}

By using the orthogonality of this form and only using the $Q$ smallest eigenvalues we form our approximation.

\begin{equation}
\begin{split}
\tilde{M}_{eig}^{-1} \equiv \sum_{q}^{Q}{\frac{1}{\lambda^{(q)}} e_{R}^{(q)} e_{L}^{(q)^T}} \\
\text{where} \,\, e_{R}^{(q)}  \cdot e_{L}^{(q')} = \delta_{q \, q'}
\end{split}
\end{equation}

The smallest eigenvalues of $M$ are used as we expect their contributions will be the greatest. By utilizing this eigenspectrum formulation, we are better able to represent the off-diagonal elements for small quark masses. Since at small quark masses the eigenvalue spectrum shows many small eigenvalues, we expect that we will get increasingly better results in a the range of $\kappa_{crit}$. 

As before we can form the trace as

\begin{equation}
\text{Tr}\left(M^{-1}\right)=\lim_{N\to\infty} \frac{1}{N} \sum_{n}^{N} z_{2_i}^{(n)} \sum_{j}{(M_{i j}^{-1}-\tilde{M}_{eig}^{-1}) z_{2_j}^{(n)}}+\text{Tr}\left(\tilde{M}_{eig}^{-1}\right)
\end{equation}

Unlike the Perturbative method, however, the $\text{Tr}\left(\tilde{M}_{eig}^{-1}\right)$ can be formed with extreme ease. 

\begin{equation}
\text{Tr}\left(\tilde{M}_{eig}^{-1}\right) = \sum_{q}{\frac{1}{\lambda^{(q)}}}
\end{equation}

The ease of this method does not end here. Since no part of $\tilde{M}_{eig}$ is ever actually formed, but rather the right/left eigenvectors are, the number of matrix/vector multiplications is drastically minimized when compared to the Perturbative Expansion Method. 

\section{Perturbative Subtraction and Eigenspectrum Subtraction}
In an attempt to combine both method in hopes to yield a ``hybrid'' method we developed something new.  To n\"aively apply both methods at once would form an approximation matrix of the form

\begin{equation}
\tilde{M}^{-1}_{naive} = \tilde{M}^{-1}_{pert} + \tilde{M}^{-1}_{eig}.
\end{equation}

The issue with this naive formulation is the the approximate off-diagonal information would be removed twice over, introducing error. In order to correct for this issue we need to remove the low eigenspectrum information from $\tilde{M}^{-1}_{pert}$ and replace it with the low eigenspectrum information of $M^{-1}$. This new formulation would provide a perturbative approximation, but where the low eigenmode information of the approximate perturbative matrix is replaced with the low eigenmode information of $M$. This should allow us to approximate for both large and small values of $\kappa$. To subtract the low eigenmode information from $\tilde{M}^{-1}_{pert}$ we form it as

\begin{equation}
\begin{split}
\tilde{M}^{-1}_{pert-eig} \equiv \tilde{M}_{pert}^{-1} - \sum_{q}^{Q}{\frac{1}{\eta^{(q)}} e_{R}^{(q)} e_{L}^{(q)^T}} \\
\text{where} \,\, \frac{1}{\eta^{(q)}} = e_{L}^{(q)} \tilde{M}^{-1}_{pert} e_{R}^{(q)}
\end{split}
\end{equation}

$\eta$ is an eigenvalue-like term that approximates the eigenvalues of $\tilde{M}^{-1}_{pert}$. With this formulation our new trace comes to

\begin{equation}
\begin{split}
\text{Tr}(M^{-1})&=\lim_{N\to\infty} \frac{1}{N} \sum_{n}^{N} z_{2_i}^{(n)} \sum_{j}{(M_{i j}^{-1}-\tilde{M}^{-1}_{pert-eig}-\tilde{M}_{eig}^{-1}) z_{2_j}^{(n)}}+\text{Tr}\left(\tilde{M}_{eig}^{-1}\right) + \text{Tr}\left( \tilde{M}^{-1}_{pert-eig} \right) \\
&=\lim_{N\to\infty} \frac{1}{N} \sum_{n}^{N} z_{2_i}^{(n)} \sum_{j}{(M_{i j}^{-1}-\tilde{M}^{-1}_{pert-eig}-\tilde{M}_{eig}^{-1}) z_{2_j}^{(n)}}+ \sum_{q}^{Q}{\frac{1}{\lambda^{(q)}}} + \text{Tr}\left( \tilde{M}^{-1}_{pert}\right) -\sum_{q}^{Q}{\frac{1}{\eta^{(q)}}} 
\end{split}
\end{equation}

\section{Larger Problems}
The Wilson action can be formed as

\begin{equation}
M = I - \kappa
\left( {\begin{array}{cc}
 0 & H_{oe}  \\
 H_{eo} & 0  \\
 \end{array} } \right)
\end{equation}

allowing for the linear system to be solved as 

\begin{equation}
\left( I - \kappa
\left( {\begin{array}{cc}
 0 & H_{oe}  \\
 H_{eo} & 0  \\
 \end{array} } \right) \right)
\left(
{\begin{array}{c}
x_o\\x_e \\
\end{array} }
\right)=
\left(
{\begin{array}{c}
b_o\\b_e \\
\end{array} }
\right)
\end{equation}

Due to the form of the matrix, a ``reduced matrix'' can be formed allowing for the problem to be cut in half. 

\begin{equation}
M_{reduced} = \frac{1}{\kappa^2}-H_{eo}H_{oe}
\end{equation}

The reduced linear system then becomes

\begin{equation}
M_{reduced} x_e = \frac{1}{\kappa^2} b_e+\frac{1}{\kappa} H_{eo} b_o
\end{equation}

and $x_o$ can then be directly computed by

\begin{equation}
x_o = b_o+\kappa H_{oe} x_e
\end{equation}

When programming for Lattice QCD, the reduced problem is more efficient. The Eigenspectrum Subtraction information can be extracted from the reduced system in the Wilson case. The eigenspectrum of the reduced matrix will be formed in the usual way.

\begin{equation}
M_{reduced} x_e^{(q)} = \hat{\lambda}^{(q)} x_e^{(q)}
\end{equation}

The eigenvalue of the reduced matrix ($\hat{\lambda}$), can be related to the eigenvalue of the full matrix ($\lambda$).

\begin{equation}
\lambda = 1 \pm \sqrt{1-\kappa^2 \hat{\lambda}}
\end{equation}

The eigenvectors, likewise, can be be related. $x_e$, the eigenvector of the reduced matrix, is also the even part of the eigenvector of the full matrix.

\begin{equation}
\begin{split}
x_e &= x_e \\
x_o &= \pm \frac{\kappa}{\sqrt{1-\kappa^2 \hat{\lambda}}} H_{oe} x_e
\end{split}
\end{equation}

Since the reduced matrix can be used to form the eigenspectrum information of the full matrix, which is readily available by linear solver algorithms with deflation, forming the full matrix is never needed. This is untrue for the Perturbative Subtraction method. The low eigenspectrum information is related to the ``negative sign'' of $\lambda$ and $x_o$.

\section{Tests and Results}
As a test, we applied the method to an $8^4$ Wilson matrix, $M$, using a parallel version of MATLAB. The $\kappa_{crit}= 0.15701$ value was determined.  

Since we are only interested in comparing size of the error bars from different methods, and thus each methods effectiveness various values of $\kappa$, we zero out all the values themselves in the figures and only show the error bars.

The trials consisted of runs where $100$ real $Z(2)$ noises were used for 20 different times, each time with a different random seed, to define error bars. We show the results after 1, 10, 20,..., 100 noises, or iterations. In order that error bars do not overlap, the results are separated from one another by a small value on the noise axis. (The non-subtracted results are located exactly at positions 1,10, 20, 30,..., 100).

The notation is as follows.

\begin{center}
\begin{tabular}{  | l | l |}
\hline
 NS & Non-Subtracted \\
 \hline
 PE & $4^{th}$ Order Perturbative Subtraction \\
 \hline 
 \textit{Q} ev & Eigenspectrum Subtraction with \textit{Q} eigenvectors \\
 \hline
 PEc+\textit{Q} ev & Perturbative Subtraction corrected with \textit{Q} eigenvectors and\\  & Eigenspectrum Subtraction with \textit{Q} eigenvectors \\
\hline
\end{tabular}
\end{center}

\begin{figure}
\centering
\subfloat[$\kappa_{crit} = 0.15701$]{\includegraphics[scale=0.22]{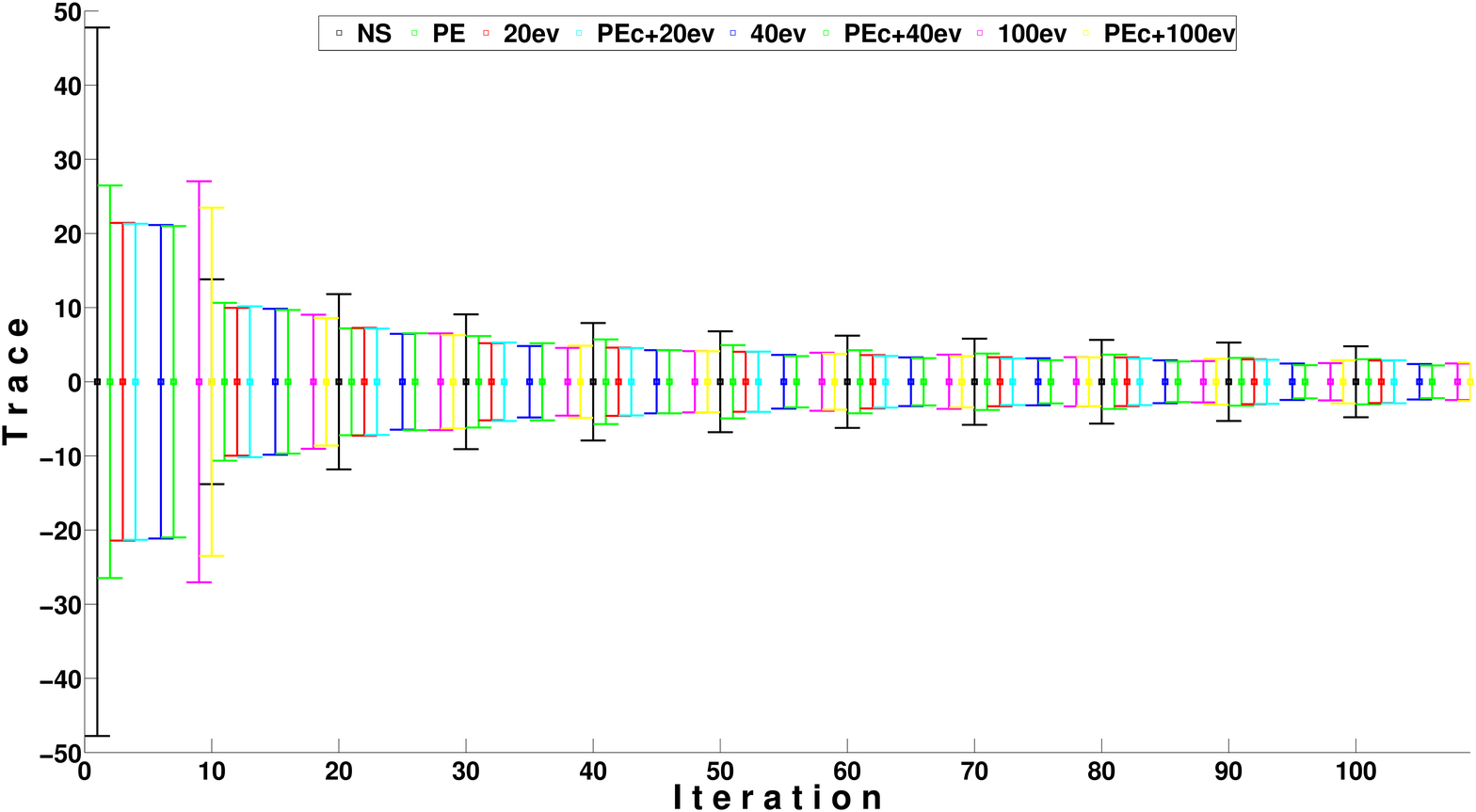}} \\
\subfloat[$\kappa = 0.1560$]{\includegraphics[scale=0.22]{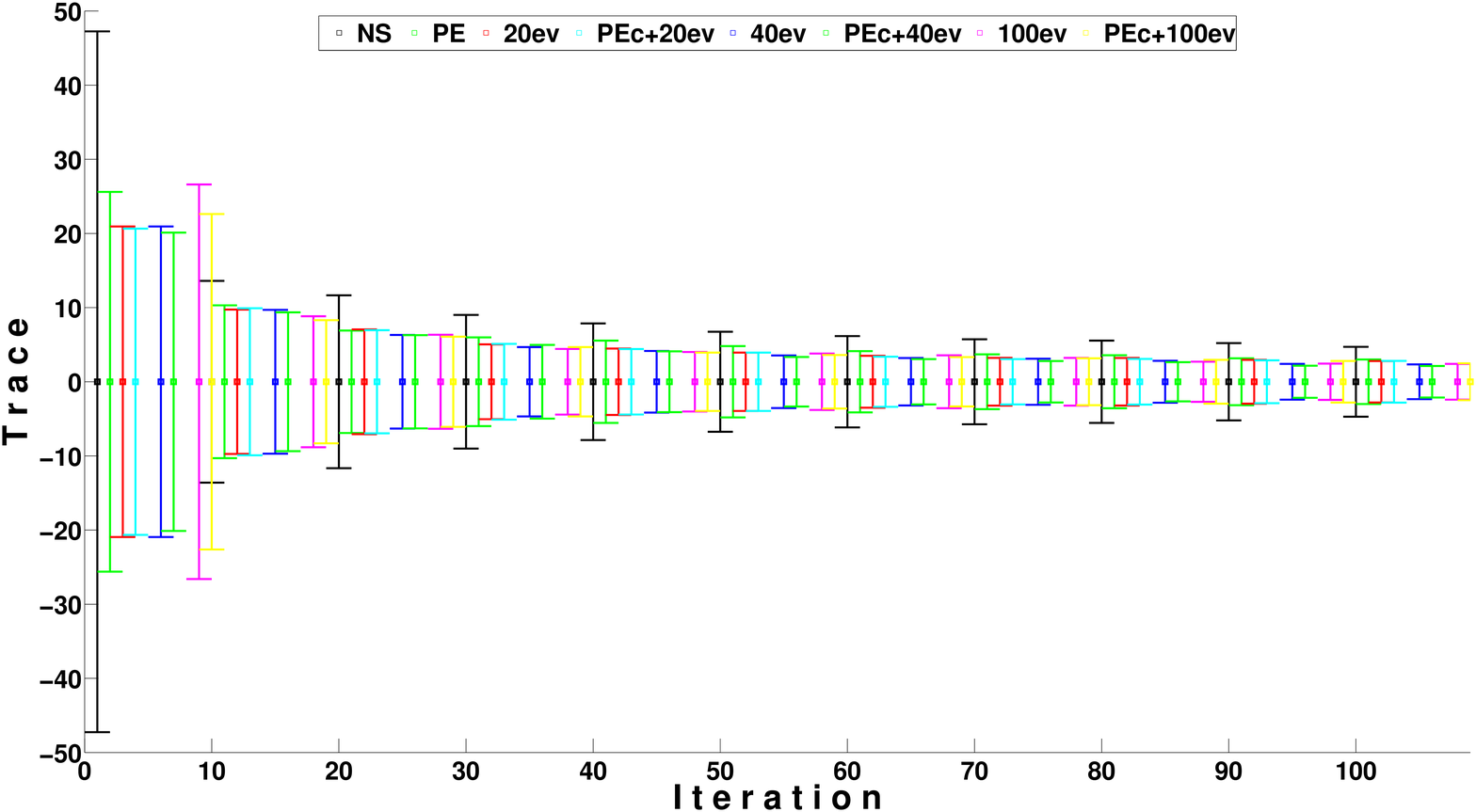}} \\
\subfloat[$\kappa = 0.1550$]{\includegraphics[scale=0.22]{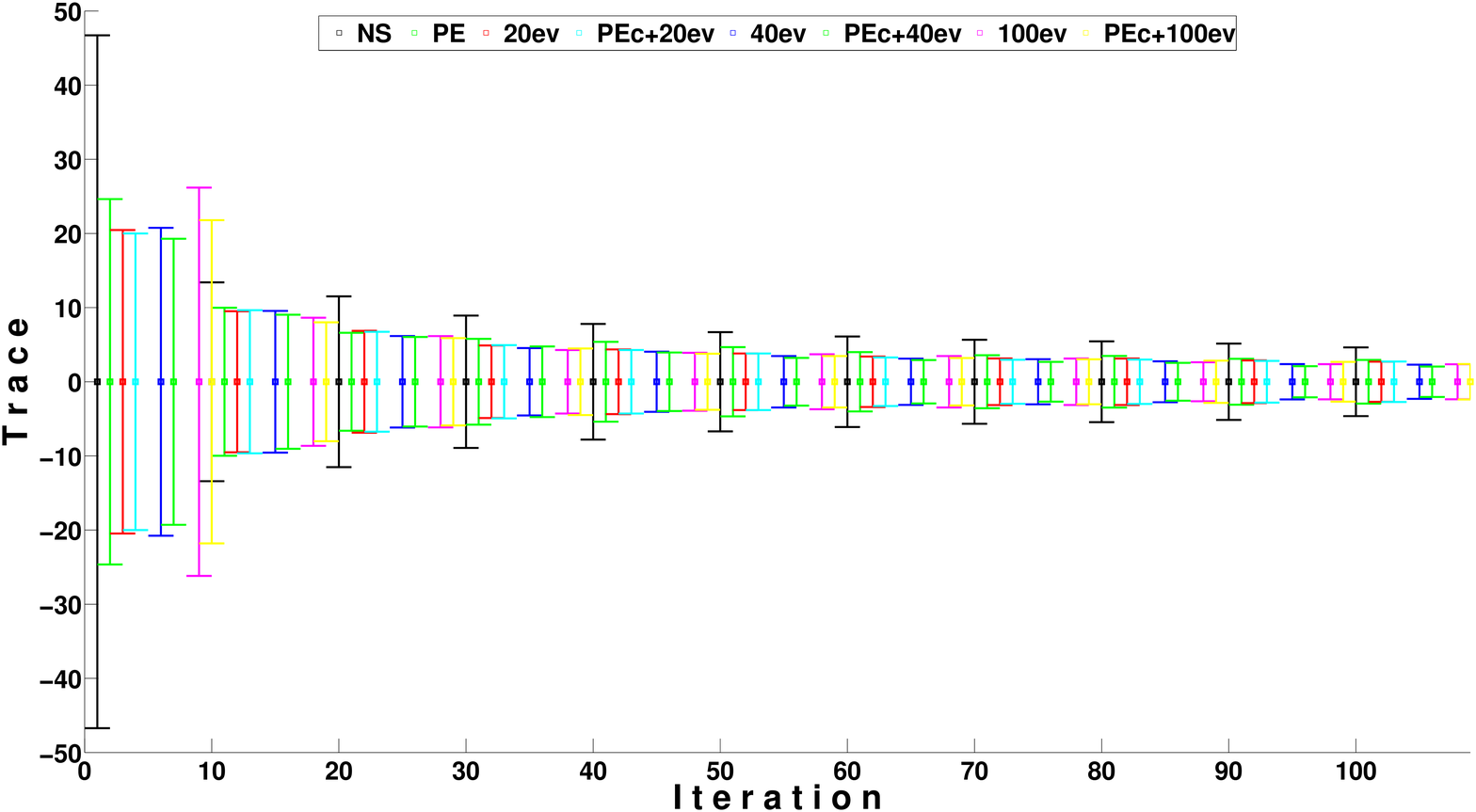}}
\caption{Comparing different levels of Eigenspectrum Subtraction, Perturbative Subtraction with Eigenspectrum Subtraction, Non-Subtracted, and $4^{th}$ Order Perturbative Subtraction.  ERROR BARS ONLY}
\end{figure}

\begin{figure}[htp]
\begin{center}
\includegraphics[scale=0.22]{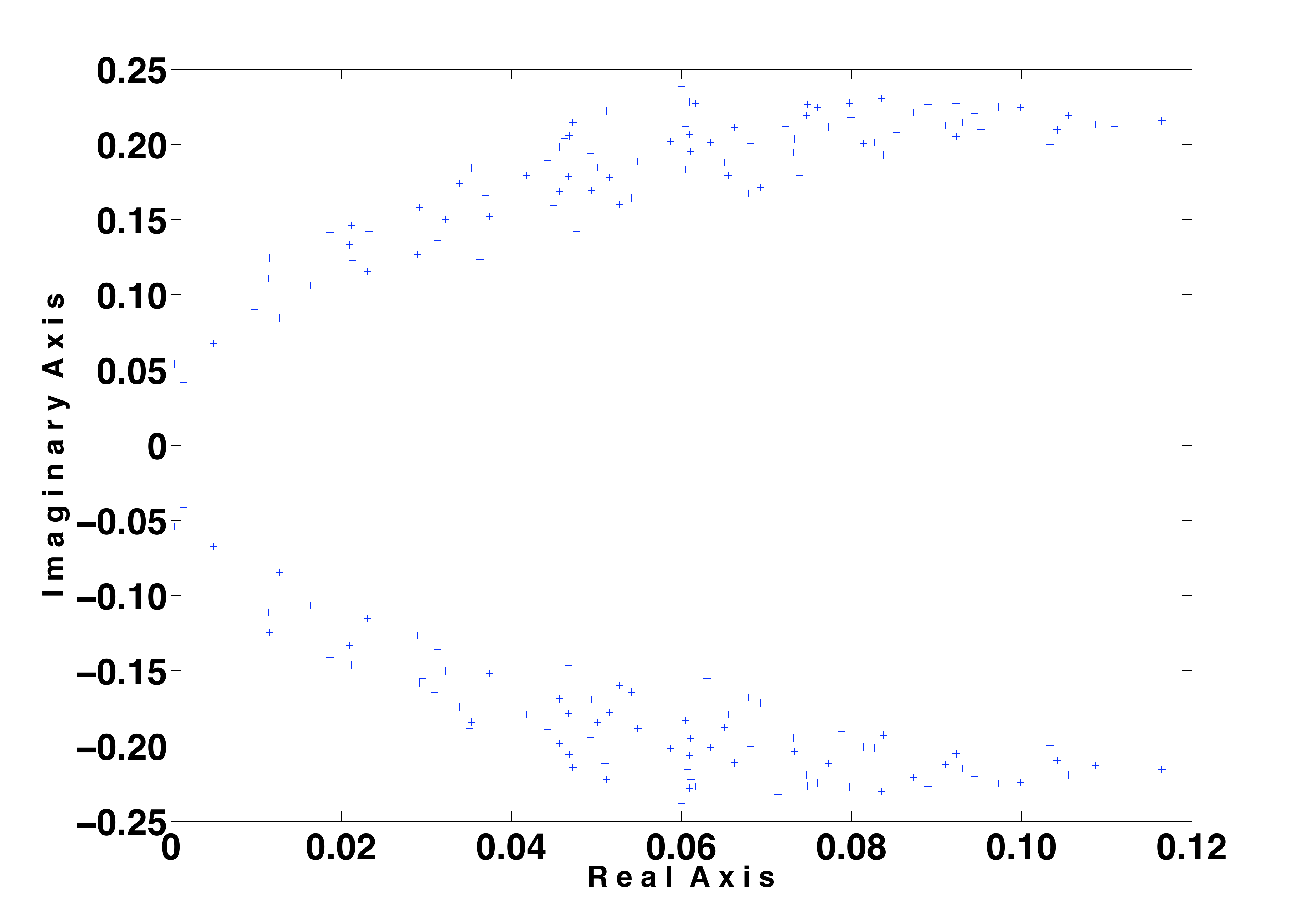}
\caption{Plot of the low eigenspectrum of the $8^4$ Wilson lattice at $\kappa_{cr}=0.15701$.}
\end{center}
\end{figure}

\section{Conclusions}

At a value of $\kappa$ near $\kappa_{crit}$ our method shows evidence of improvement over the 4th order Perturbative Subtraction Method. By combining the two methods, and removing the low eigen-information from the Perturbative Subtraction Method, we see an additional increased improvement, albeit small. As $\kappa$ deviates from $\kappa_{crit}$ we see results on par with Perturbative Subtraction.

We would expect our method to become more efficient for larger matrices with smaller eigenvalues, but we need more study. The combined effects of Perturbative Subtraction and Eigenspectrum Subtraction are still being studied as well. Development in FORTRAN is in progress to facilitate the larger matrices. 

\section{Acknowledgments}
Calculations done with HPC systems at Baylor University.

\bibliographystyle{plain}
\bibliography{sample}

\end{document}